\documentclass[a4paper,11pt]{article}
\usepackage{jinstpub} 
\usepackage{lineno}
\usepackage{orcidlink}
\usepackage{float}
\usepackage{appendix}
\usepackage{makecell}
\usepackage{graphicx}
\usepackage{amsmath}
\usepackage{url}
\usepackage{hyperref}

\title{\boldmath The SND@LHC neutron shielding}
\collaboration{The SND@LHC collaboration}
\author{D.~Abbaneo$^{1}$\orcidlink{0000-0001-9416-1742},}
\author{S.~Ahmad$^{4}$\orcidlink{0000-0001-8236-6134},}
\author{R.~Albanese$^{2,3}$\orcidlink{0000-0003-4586-8068},}
\author{A.~Alexandrov$^{2}$\orcidlink{0000-0002-1813-1485},}
\author{F.~Alicante$^{2,3}$\orcidlink{0009-0003-3240-830X},}
\author{F.~Aloschi$^{1}$\orcidlink{0000-0002-2501-7525},}
\author{K.~Androsov$^{5}$\orcidlink{0000-0003-2694-6542},}
\author{C.~Asawatangtrakuldee$^{6}$\orcidlink{0000-0003-2234-7219},}
\author{M.A.~Ayala~Torres$^{7,8}$\orcidlink{0000-0002-4296-9464},}
\author{N.~Bangaru$^{2,3}$\orcidlink{0009-0004-3074-1624},}
% ...

\affiliation{$^{1}$European Organization for Nuclear Research (CERN), Geneva, 1211, Switzerland}
\affiliation{$^{2}$Sezione INFN di Napoli, Napoli, 80126, Italy}
\affiliation{$^{3}$Universit\`{a} di Napoli ``Federico II'', Napoli, 80126, Italy}
\affiliation{$^{4}$Affiliated with Pakistan Institute of Nuclear Science and Technology(PINSTECH), Nilore, 45650, Islamabad, Pakistan}
\affiliation{$^{5}$Institute of Physics, EPFL, Lausanne, 1015, Switzerland}
\affiliation{$^{6}$Chulalongkorn University, Bangkok, 10330, Thailand}
\affiliation{$^{7}$Center for Theoretical and Experimental Particle Physics, Facultad de Ciencias Exactas, Universidad Andr\`es Bello, Fernandez Concha 700, Santiago, Chile}
\affiliation{$^{8}$Millennium Institute for Subatomic physics at high energy frontier-SAPHIR, Santiago, 7591538, Chile}
% ...

\emailAdd{jilberto.zamora@unab.cl}

\begin{document}

\maketitle

\begin{abstract}

This paper presents the design, construction, and simulation-based validation of the ColdBox, a combined neutron shielding and insulating enclosure for the Scattering and Neutrino Detector at the LHC (SND@LHC). The emulsion films in the detector's target region require protection from the intense neutron radiation background and a stable environment of \(15 \pm 1\)\,\(^\circ\)C and 50--55\,\% relative humidity for long-term stability. The ColdBox meets these requirements through a dual-layer structure: an external 5\,cm plexiglass wall to moderate fast neutrons, and an internal 4\,cm layer of borated polyethylene (with 35\,\% boron content) to capture thermal neutrons. The mechanical design, based on a robust aluminum frame, accommodates the constraints of the TI18 tunnel. FLUKA simulations were used to optimize the shielding configuration, showing a significant reduction in the neutron flux, with a simulated ratio of shielded to unshielded thermal neutron fluence of \(2.3 \times 10^{-3}\). This result is consistent with initial measurements from BatMon detectors. The design also provides a sealed volume for a cooling system to maintain the required temperature and humidity, ensuring the necessary conditions for the emulsion films' integrity.
\end{abstract}

\section{Introduction}
\label{sec:introduction}

The Scattering and Neutrino Detector at the Large Hadron Collider (SND@LHC)~\cite{SNDLHC:2022ihg} is a compact hybrid apparatus located in the TI18 tunnel, 480 meters downstream from the ATLAS interaction point (IP1). It is designed to detect and distinguish the three flavors of neutrinos (\(\nu_{e}\), \(\nu_{\mu}\), and \(\nu_{\tau}\)) in the high-energy regime (100 GeV - 1 TeV) within the pseudo-rapidity range of \(7.2 < \eta < 8.6\).

\begin{figure}[H] 
\centering
\includegraphics[scale=0.25]{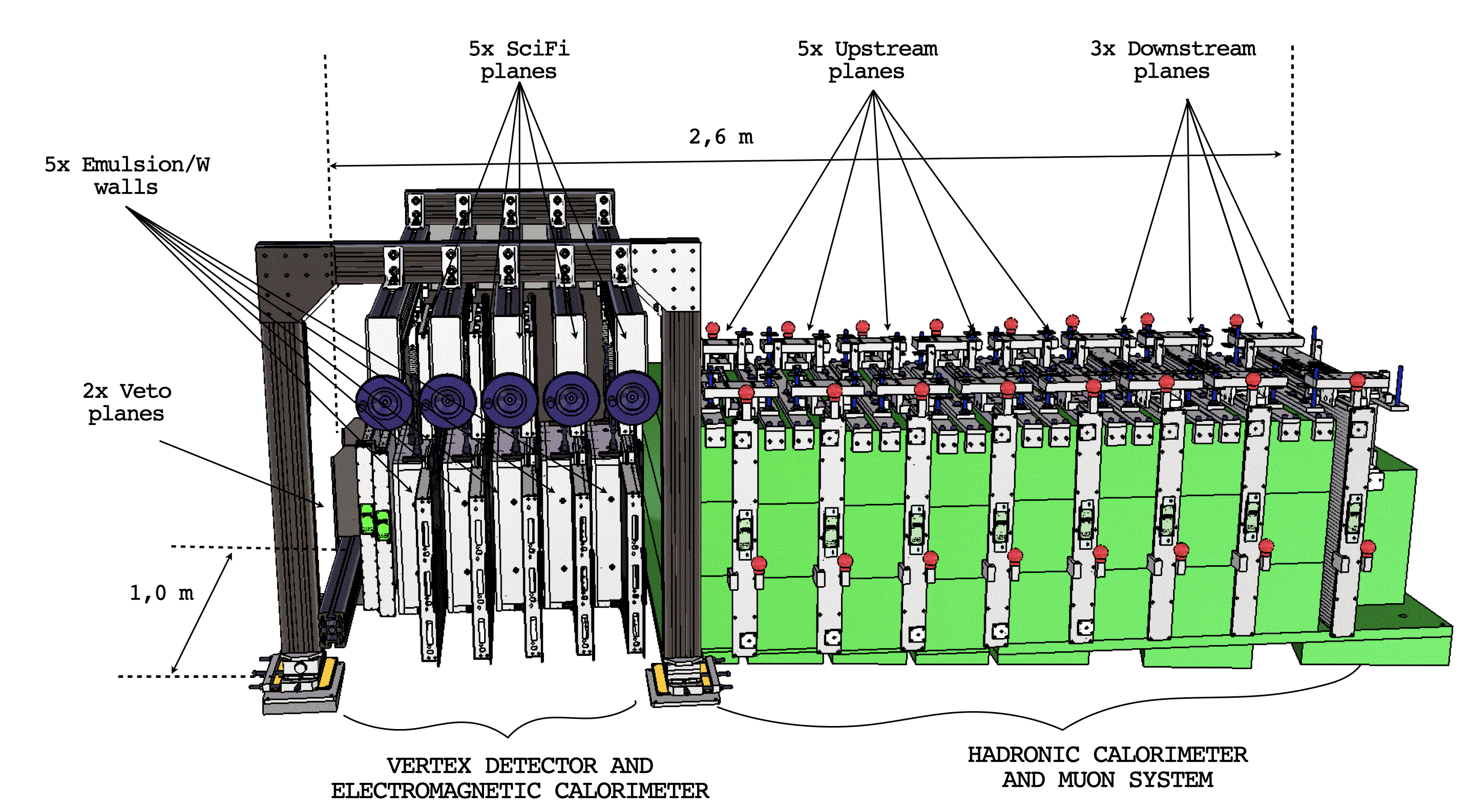}
\caption{The SND@LHC detector.}
\label{snd_detectors}
\end{figure}

The apparatus, shown in Figure~\ref{snd_detectors}, comprises three main subsystems. The upstream veto system identifies incoming muons from IP1. The central target region---defined for this paper as the vertex detector combined with the electromagnetic calorimeter---contains emulsion films interleaved with tungsten plates for neutrino interaction detection. The downstream muon system---defined as the hadronic calorimeter plus the muon identification system---acts as a coarse sampling calorimeter.

The emulsion films in the target region are highly sensitive to environmental conditions and radiation. For long-term stability, they require a strictly controlled environment of \(15 \pm 1\)\,\(^\circ\)C and 50--55\,\% relative humidity. Furthermore, they must be shielded from the intense neutron background present in the tunnel. This paper details the design, construction, and performance of a single structure, the "ColdBox", which meets the primary requirements of a sealed environment and neutron shielding. Furthermore, the design is future-proofed with ample insulation to accommodate a cooling system, the integration and performance of which fall outside the scope of the present study.

The mechanical design was constrained by the compact and curved TI18 tunnel, which features a ventilation pipe and cable trays along the wall adjacent to the detector. The design had to ensure a minimum transit aisle, provide access for an emulsion replacement trolley, and conform to the tunnel's sloped floor. The chosen shielding strategy employs a dual-layer approach: an external moderator of plexiglass to thermalize fast neutrons, followed by an internal absorber of borated polyethylene to capture them.

This report is structured as follows: Section~\ref{sec:mechanical} describes the mechanical design, construction, and structural analysis of the ColdBox. Section~\ref{sec:fluka} presents the FLUKA simulations used to optimize the shielding configuration and assesses its performance. Section~\ref{sec:verification} verifies the material composition of the borated polyethylene. Our conclusions are given in Section~\ref{sec:conclusions}.

\section{Mechanical Design and Construction}
\label{sec:mechanical}

\subsection{Design Requirements and Frame}

\begin{figure}[H]
\centering
\includegraphics[scale=0.055]{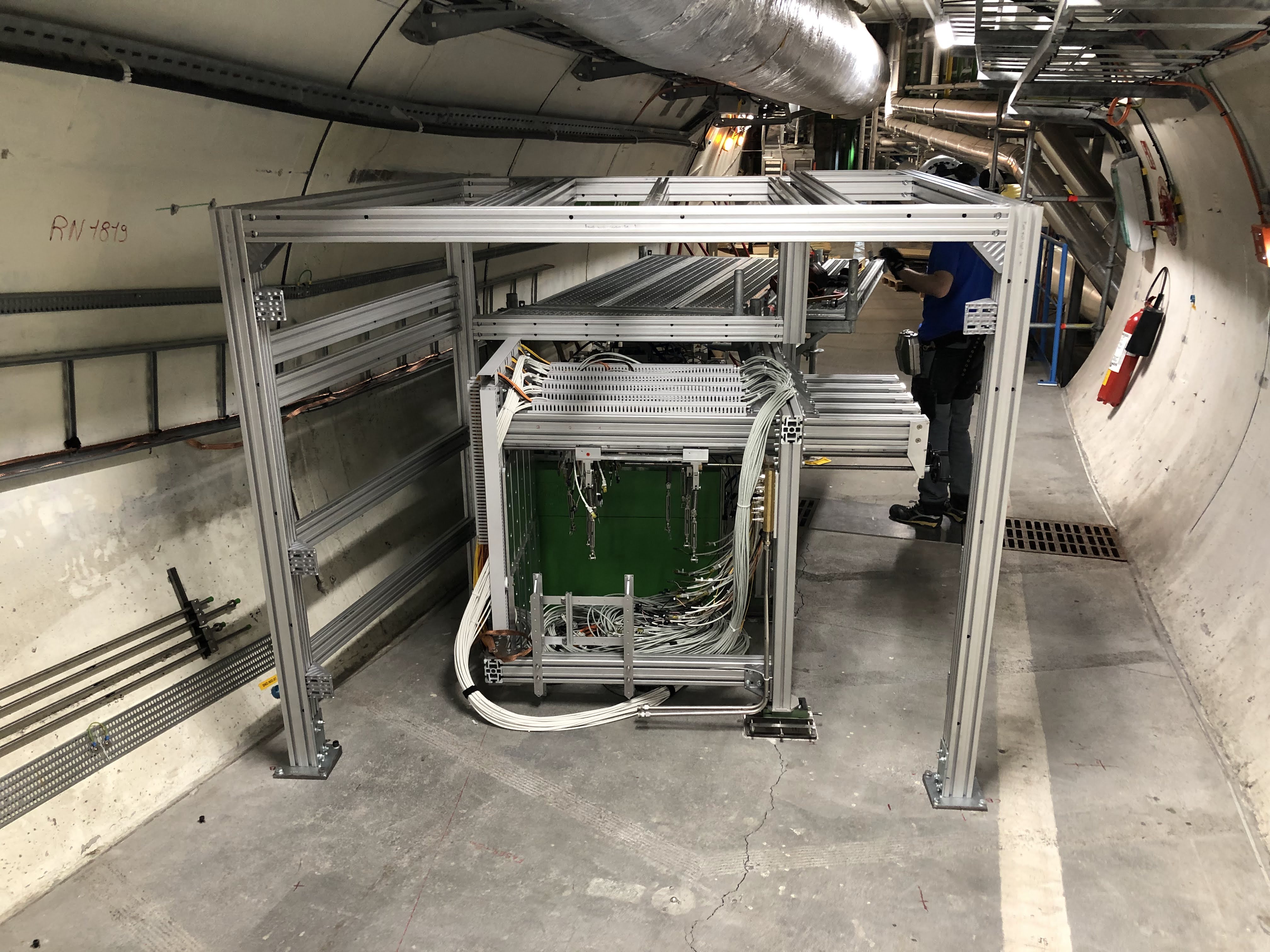}
\caption{Installed aluminium frame}
\label{aluframe}
\end{figure}

\begin{figure}[H]
\centering
\includegraphics[scale=0.35]{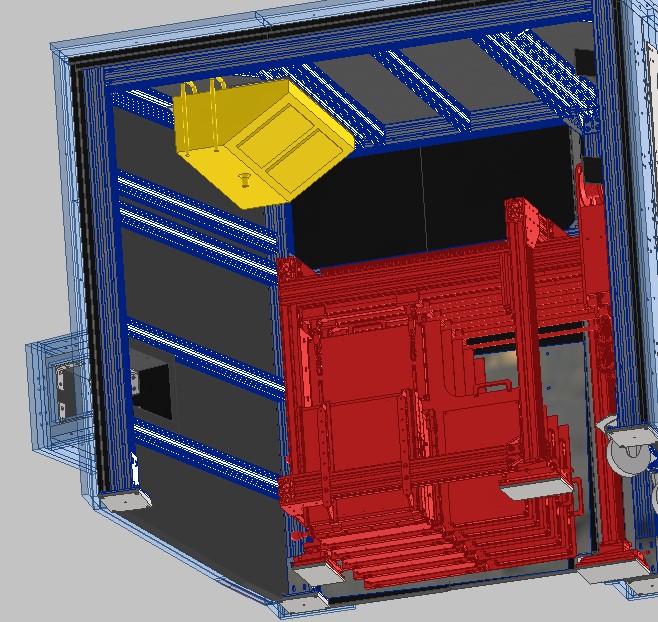}
\caption{Evaporator highlighted in yellow over the target region highlighted in red.}
\label{evaporator}
\end{figure}

The target region has an internal structure made of aluminium profiles, anchored to the floor via three feet. A primary design condition was that the ColdBox must not contact this internal structure to avoid imparting stress or movement to the detectors. The ColdBox therefore features its own self-supporting internal frame (Figure~\ref{aluframe}) constructed from T-slotted aluminium profiles (specifications in Table~\ref{tab:aluframe}), designed to support the estimated 400\,kg of shielding materials and four doors, each weighing approximately 150\,kg. The frame was also designed to provide the necessary clearance for the emulsion replacement trolley and to host an evaporator unit on its roof (Figure~\ref{evaporator}) to maintain the required internal climate within the $\sim$5\,m$^3$ volume.

\begin{table}[h!]
\centering
  \begin{tabular}{| l | c |}
    \hline
    Company & Bosch Rexroth \\ \hline
    Material & Aluminium \\ \hline
    Profile measures 1 & 80x80 $\rm mm^2$ \\ \hline
    Profile measures 2 & 40x80 $\rm mm^2$ \\ \hline
    Density & 2.7 ${\rm g}/{\rm cm^3}$ \\ \hline
    Melting point & 660 $\rm ^{o}$C \\
    \hline
  \end{tabular}
\caption{Details of the aluminium frame}
\label{tab:aluframe}
\end{table}

\subsection{Shielding Materials and Assembly}

The shielding walls consist of an external 50\,mm layer of plexiglass and an internal 40\,mm layer of borated polyethylene. Material specifications are provided in Tables~\ref{tab:borated} and~\ref{tab:plexiglass}. 

\begin{table}[h!]
\centering
\begin{minipage}[b]{0.48\textwidth}
\centering
  \begin{tabular}{| l | c |}
    \hline
    Company & \makecell{Tangyin Sanyou \\ Engineering} \\ \hline
    Material & Borated polyethylene \\ \hline
    Boron-carbide content & $30\%$ \\ \hline
    Thickness & 40 mm \\ \hline
    Density & $\rm 1.2 {g}/{cm^3}$ \\
    \hline
  \end{tabular}
  \caption{Borated polyethylene information provided by the manufacturer.}
  \label{tab:borated}
\end{minipage}
\hfill
\begin{minipage}[b]{0.48\textwidth}
\centering
  \begin{tabular}{| l | c |}
    \hline
    Company & \makecell{Chongqing Niubai \\ Electromechanical} \\ \hline
    Material & Plexiglass \\ \hline
    Thickness & 50 mm \\ \hline
    Density & 1.19 $\rm {g}/{cm^3}$ \\ \hline
    Melting point & 175 $\rm ^{o}C$ \\ \hline
    Flammable & 375 $\rm ^{o}C$  \\
    \hline
  \end{tabular}
  \caption{Plexiglass information provided by the manufacturer.}
  \label{tab:plexiglass}
\end{minipage}
\end{table}

\begin{figure}[h!]
\centering
\includegraphics[scale=0.1]{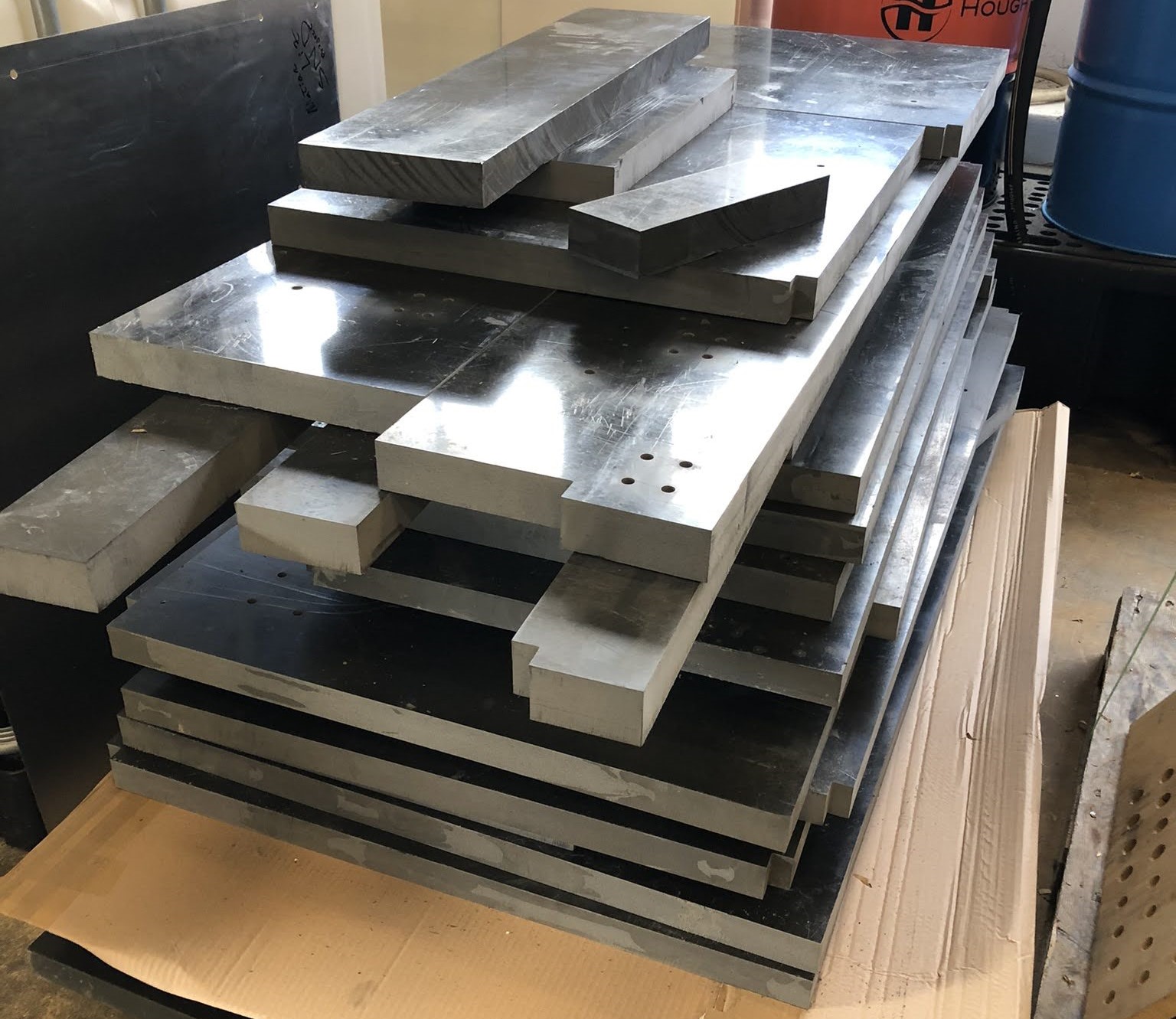} \hspace{0.5cm}
\includegraphics[scale=0.052]{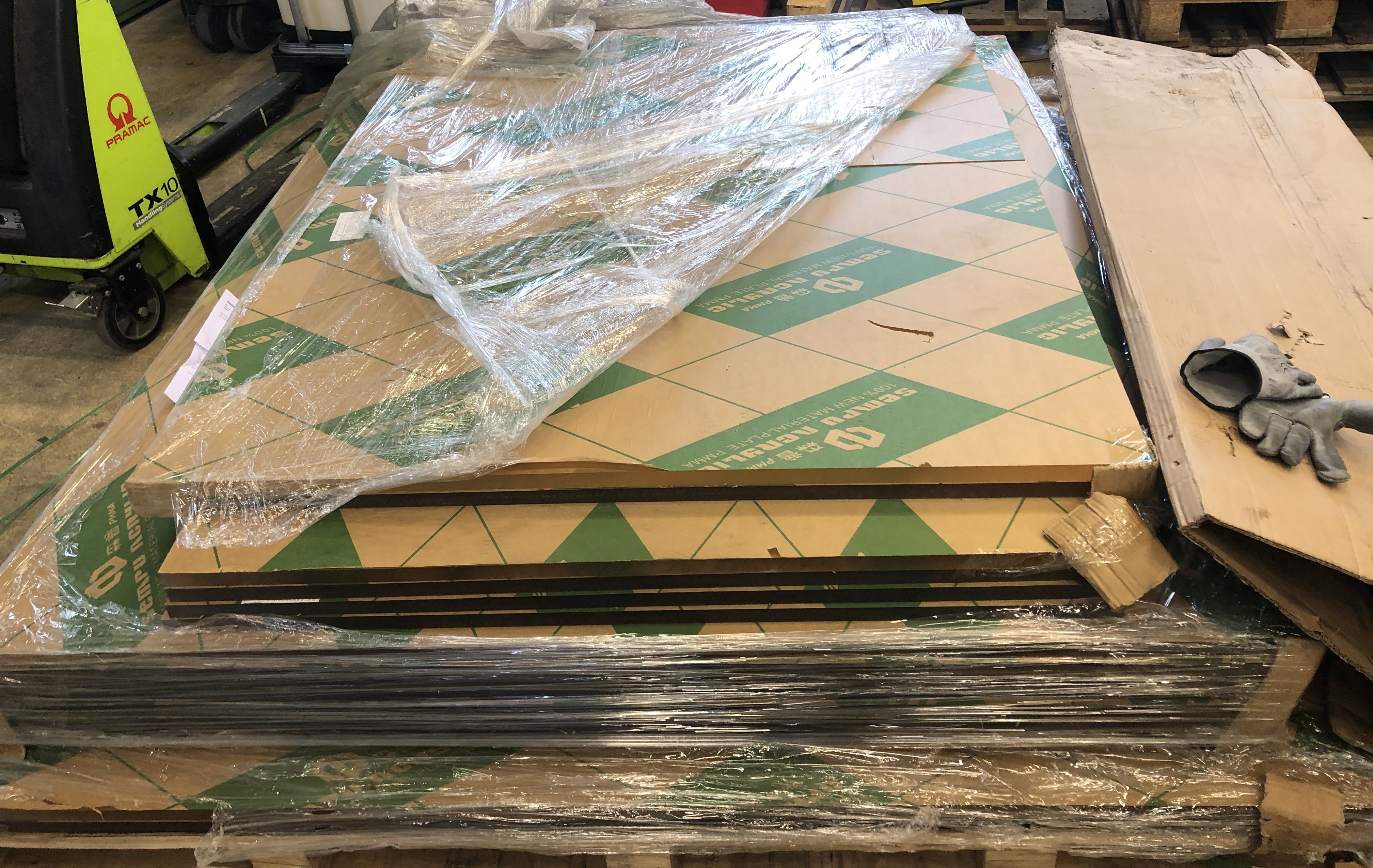}
\caption{Left side: machined borated polyethylene. Right side: plexiglass sheets before the machining process.}
\label{polplex}
\end{figure}

The borated polyethylene (see Figure~\ref{polplex}, left side)  behaves as a thermo-stable plastic similar to High-Density Polyethylene (HDPE). The two applied solutions to machine this material were water-jet cutting for rectangular contour shapes plus holes, and SPHW1204PDR-A88 WCD10 Polycrystalline Diamond (PCD) tool insert with 1.8 $\rm m/min$ of feed in a milling machine for angled contour roughing and surface roughing. 

Plexiglass behaves as a thermoplastic that can be manufactured under regular conditions. Due to the significant size and required thickness, ordering two layers of 25 mm (see Figure~\ref{polplex}, right side) was necessary to reach a total of  50 mm thickness. To obtain rectangular contour shapes, the cutting method applied for this material was to use an industrial saw machine. It was also necessary to apply the drilling process for holes and angled contour roughing.

\begin{figure}[h!]
\centering
\includegraphics[scale=0.16]{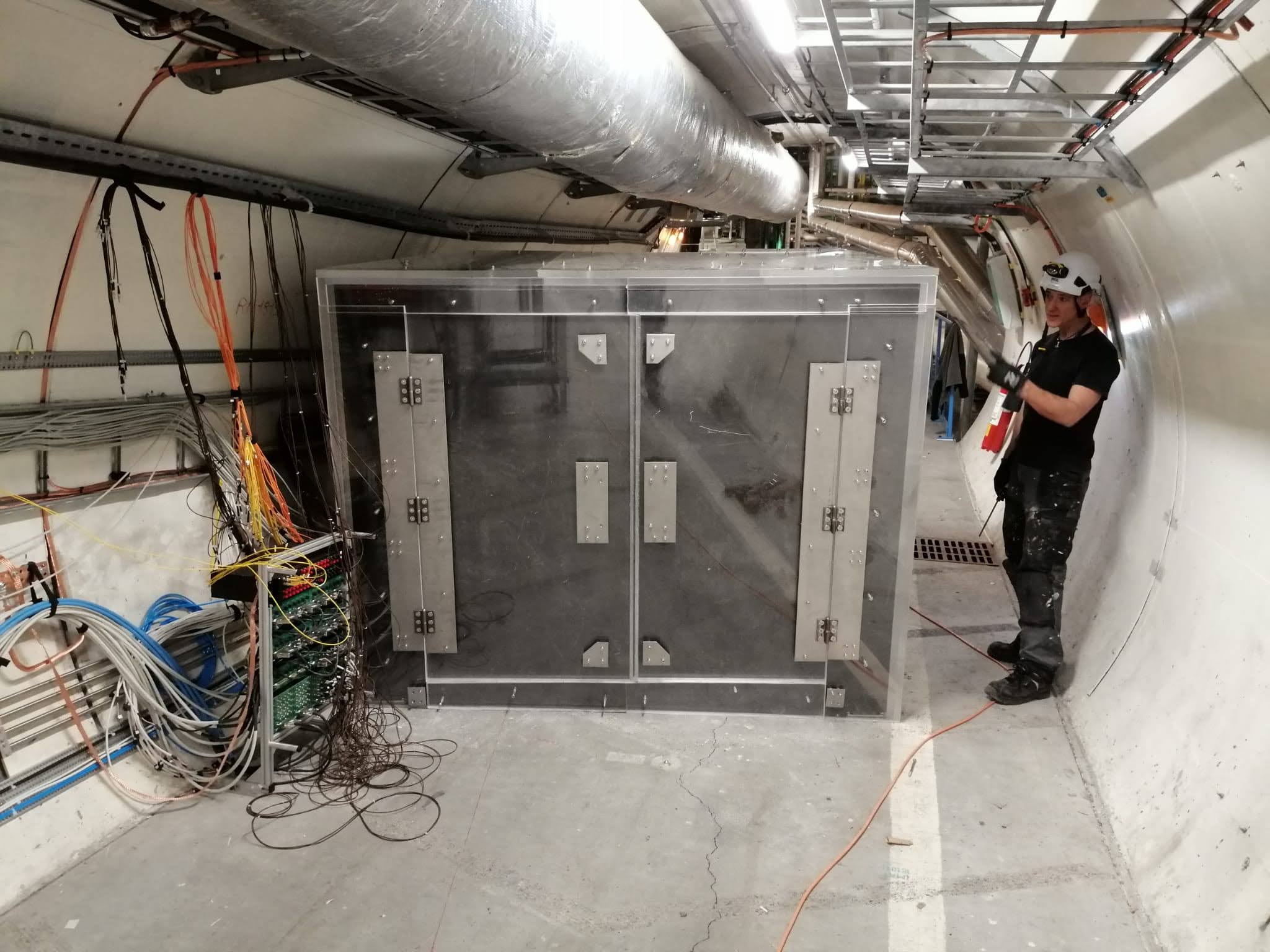}
\caption{View of the final ColdBox.}
\label{finalcoldbox}
\end{figure}

The final assembly, with overall dimensions of $2190 \times 1780 \times 1864$\,mm$^3$, is shown in Figure~\ref{finalcoldbox}. One wall is angled at 15$^\circ$ to conform to the TI18 tunnel curvature. Technical drawings are provided in the Appendix~\ref{appendix}.

\subsection{Sealing and Safety Compliance}

\begin{figure}[H]
\centering
\includegraphics[scale=0.098]{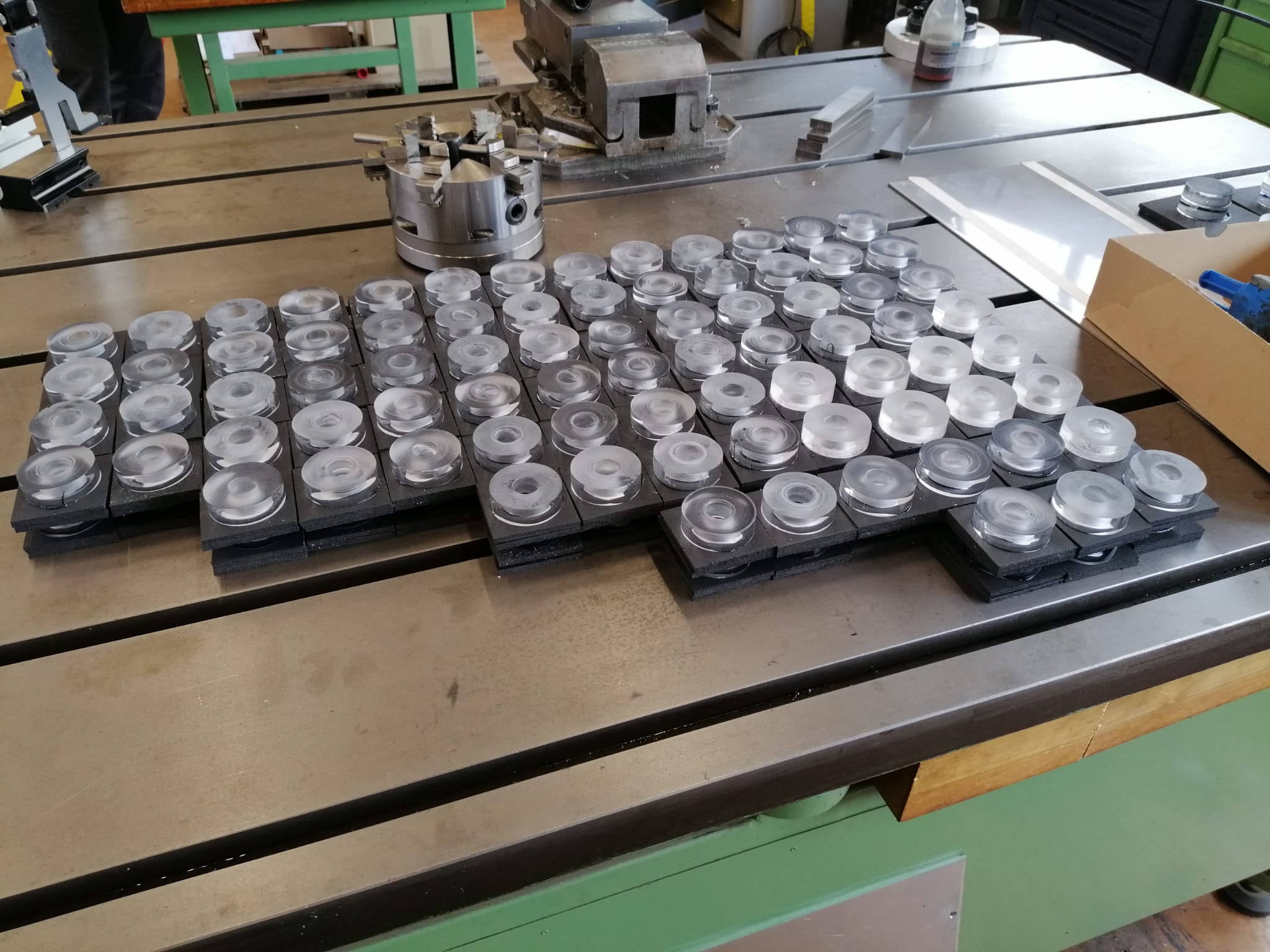} \hspace{5mm}
\includegraphics[scale=0.05]{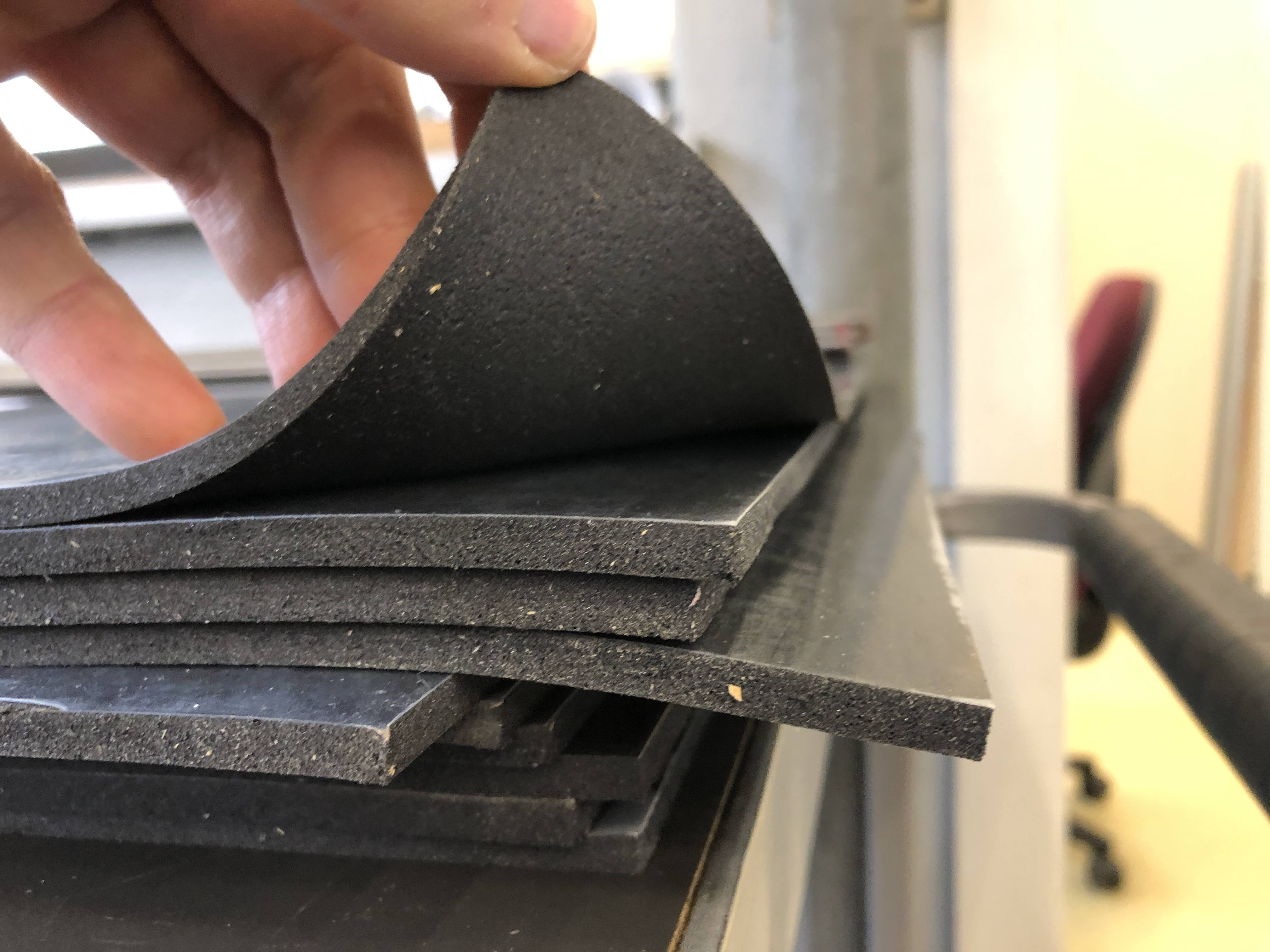}
\caption{Left side: Mirrobor seals prepared to cover bolts and screws holes in the ColdBox. Right side: Mirrobor sheets before the machining process.}
\label{mirrobor}
\end{figure}

\begin{figure}[H]
\centering
\includegraphics[scale=0.4]{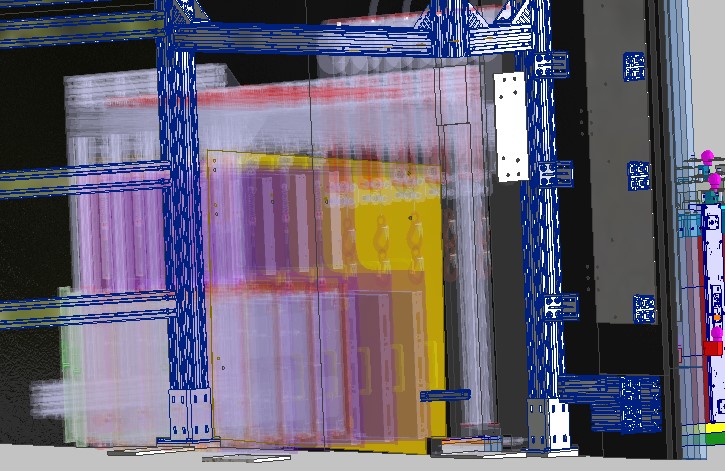}
\caption{Aluminium sheet highlighted in yellow.}
\label{alumsheet}
\end{figure}

Holes for cables, pipes, and mechanical fasteners were covered with 5\,mm thick Mirrobor sheets (Figure~\ref{mirrobor} left side, specifications in Table~\ref{tab:mirrobor}), a flexible, boron-rich material that maintains shielding integrity (see Figure~\ref{mirrobor} right side). In accordance with CERN Safety Instruction IS41 \cite{IS41}, which covers the use of plastics and other non-metallic materials for fire safety and radiation hardness, it was an issue to detail the specific parameters for possible ignition points for materials such as borated polyethylene because the manufacturer did not provide a proper datasheet with certified tests. For this reason, it was necessary to add an aluminium sheet of 1030x815x1 $\rm mm^3$ (see Figure~\ref{alumsheet}) between the target region and the muon system to prevent excess heat transfer between detectors. This solution was reviewed and approved by the CERN safety team.

\begin{table}[h!]
\centering
  \begin{tabular}{| l | c |}
    \hline
    Company & Mirrotron \\ \hline
    Material & Mirrobor \\ \hline
    Boron-carbide content & 80 $\%$ \\ \hline
    Thickness & 5 mm \\ \hline
    Density & 1.36 $\rm {g}/{cm^3}$ \\ \hline
    Melting point & 175 $\rm^{o}$C \\ \hline
    Flammable & 375 $\rm ^{o}$C  \\
    \hline
  \end{tabular}
\caption{Mirrobor information provided by the manufacturer.}
\label{tab:mirrobor}
\end{table}

\subsection{Structural Analysis}

Finite element analysis using ANSYS 2019 R1 \cite{ANSYS2019} was performed to verify the structural integrity of the aluminium frame under two load scenarios: with all doors closed and with all doors open. The results, summarized in Tables~\ref{Table5} and~\ref{Table6}, confirm that the maximum von Mises stress and deformation are within safe limits for aluminium, with the door hinge regions identified as the most critically stressed areas.

Mirrobor (see Figure~\ref{mirrobor} right side) is a rubber-like flexible material with a high boron concentration, developed by the Mirrotron manufacturer. It can be acquired in two thicknesses, 2 or 5 mm, is easy to cut by hand (with just a cutter), and is characterized as a stable, non-toxic, non-hazardous material for transport. However, high humidity or contact with acidic/organic solvent materials must be avoided when stored. Mirrobor was used to cover holes in the plexiglass walls.

\begin{table}[h!]
\centering
\begin{minipage}{0.45\textwidth}
\centering
  \begin{tabular}{| l | c |}
    \hline
    Profile & 80x80 mm \\ \hline
    Material & Aluminium \\ \hline
    Overall Von Mises stress & 5 to 15 $\rm Mpa$ \\ \hline
    Maximum stress & 40 $\rm Mpa$ \\ \hline
    Maximum deformation & 0.4 mm \\ 
    \hline
  \end{tabular}
  \caption{Finite element method analysis for closed doors scenario.}
  \label{Table5}
\end{minipage}
\hfill
\begin{minipage}{0.45\textwidth}
\centering
  \begin{tabular}{| l | c |}
    \hline
    Profile & 80x80 mm \\ \hline
    Material & Aluminium \\ \hline
    Overall Von Mises stress & 5 to 10 $\rm Mpa$ \\ \hline
    Maximum stress & 86 $\rm Mpa$ \\ \hline
    Maximum deformation & 1.1 mm \\ 
    \hline
  \end{tabular}
  \caption{Finite element method analysis for opened doors scenario.}
  \label{Table6}
\end{minipage}
\end{table}

The final design of the ColdBox has the maximum dimensions of 2190x1780x1864 $\rm mm^3$ (Figure~\ref{finalcoldbox}). While the roof and the two walls that contain doors are flat and squared with frames that follow the slope of the floor. The separation between the target region and the muon system is an escalated profile, and the parallel wall that follows the TI-18 tunnel has an angle of 15$^{o}$ compared to the internal frame of the coldbox. Detailed drawings can be found in the appendix~\ref{appendix}.

\section{FLUKA Simulations and Shielding Performance}
\label{sec:fluka}

\subsection{Simulation Setup and Material Selection}

\begin{figure}[h!]
\centering
\includegraphics[scale=0.4]{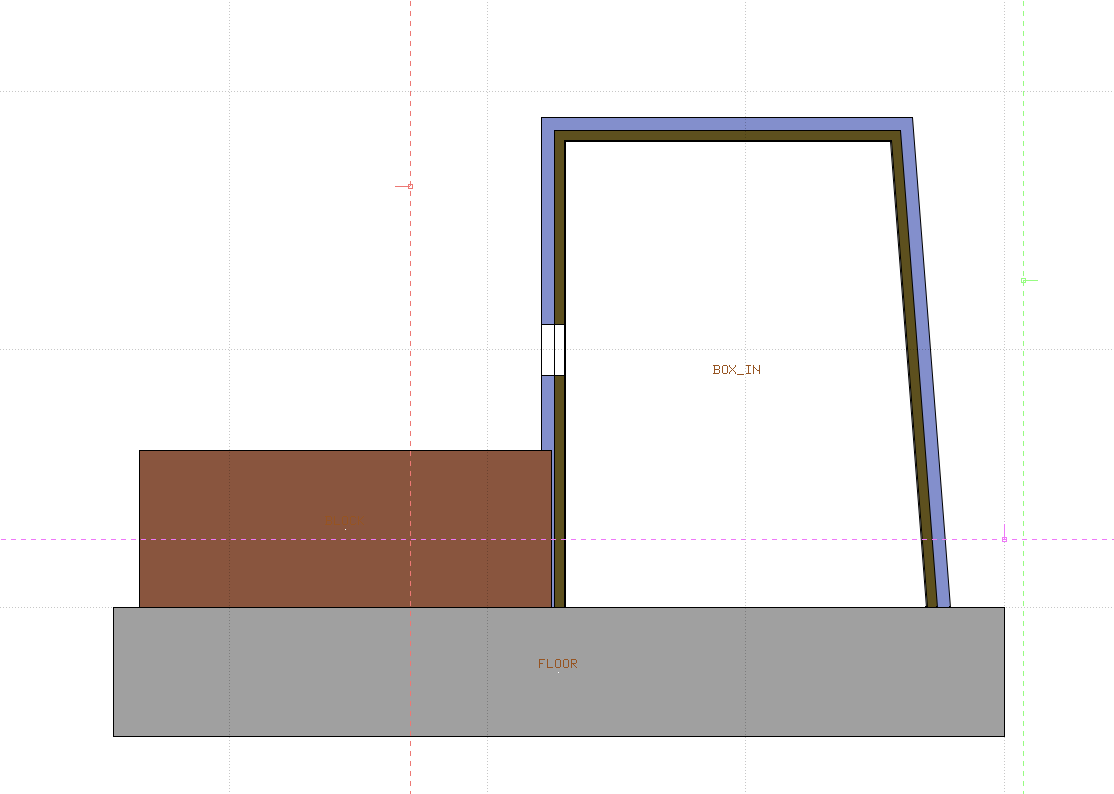}
\caption{Lateral view of the ColdBox. The brown region represents the Muon System, the grey area represents the concrete base, the blue region represents plexiglass, and the olive region represents borated polyethylene.}
\label{coldbox1}
\end{figure}

The shielding configuration was optimized using FLUKA simulations~\cite{Ahdida:2022gjl,Battistoni:2015epi} with the Flair interface~\cite{Donadon:2024omp}. The simulated geometry consisted of a $2 \times 2 \times 2$\,m$^3$ ColdBox placed on a 0.5\,m thick concrete base, with the muon system represented by an iron block (Figure~\ref{coldbox1}). The neutron source distribution was modeled based on one year of High-Luminosity LHC (HL-LHC) operation with an integrated luminosity of 250\,fb$^{-1}$.

Several material combinations were explored. For the moderator, an external layer of 5 cm plexiglass was selected over high-density polyethylene due to its superior mechanical strength\footnote{Tensile Strength limit is 64.8 - 83.4 MPa for plexiglass, while for polyethylene is 11.0 - 43.0 MPa}. For the absorber, an internal layer of 4 cm borated polyethylene with 5\%, 30\%, and 35\% boron content was evaluated; cadmium was excluded for safety reasons due to its toxicity. The 35\% boron content was selected for its highest neutron capture cross-section.

\subsection{Configuration Optimization and Results}

\begin{figure}[h!]
\centering
\includegraphics[scale=1.2]{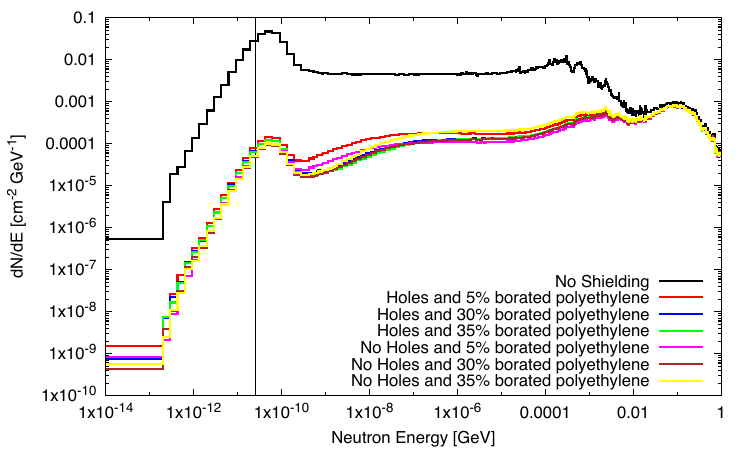}
\caption{Neutron spectrum inside the ColdBox volume. Here $10^{9}$ primary neutrons were simulated. The black vertical line at $\rm E= 0.025$ (eV) stands for thermal neutrons at $\rm 20^{o}$ C.}
\label{neutron_spectrum}
\end{figure}

The neutron fluence inside the ColdBox was calculated for different wall configurations (Figure~\ref{neutron_spectrum}):

\begin{itemize}
\item \textbf{With Holes:} Configurations with 5\%, 30\%, and 35\% boron content, and holes modeling the presence of apertures for cables and pipes.
\item \textbf{Without Holes:} Idealized configurations with 5\%, 30\%, and 35\% boron content, no holes were considered to isolate the effect of apertures for cables and pipes.
\end{itemize}

Figure~\ref{neutron_spectrum} demonstrates that the chosen ``Holes and 35\% borated polyethylene'' configuration is highly effective. The effect of the holes is negligible for thermal neutrons. The simulated ratio of neutron fluence inside the shielded ColdBox to the unshielded case, $R_{\text{SIM}}$, is defined as:
\begin{equation}
R_{\text{SIM}} = \frac{\text{Fluence with shielding (including holes)}}{\text{Fluence without shielding}}.
\label{ratio_sim}
\end{equation}
For thermal neutrons (0.025\,eV), $R_{\text{SIM}} = 2.3 \times 10^{-3}$, as shown in Table~\ref{ratio}.

\begin{table}[h!]
\centering
\begin{tabular}{|l|l|}
\hline
        Neutron Energy & $R_{\rm SIM}$ \\ \hline
        $=$ 0.025 $\rm eV$ & 2.3 x $10^{-3}$ \\ \hline 
        $<$ 1 $\rm eV$ & 2.9 x $10^{-3}$ \\ \hline
        $<$ 100 $\rm eV$ & 1.6 x $10^{-2}$ \\ \hline
        $<$ 10 $\rm keV$ & 2.8 x $10^{-2}$ \\ \hline
        $<$ 2 $\rm MeV$ & 8.0 x $10^{-2}$ \\ \hline
        $<$ 20 $\rm MeV$ & 3.0 x $10^{-1}$  \\ \hline
        $<$ 200 $\rm MeV$ & 7.3 x $10^{-1}$ \\ \hline
        $<$ 700 $\rm MeV$ & 7.0 x $10^{-1}$ \\ \hline  
\end{tabular}
\caption{Ratio $R_{\rm SIM}$}
\label{ratio}
\end{table}

\subsection{Comparison with Measurements}

The simulation results were compared with in-situ measurements from BatMon neutron monitors~\cite{BatMon1,BatMon2}. The measured ratio, $R_{\text{BatMon}}$, of thermal neutron fluence inside the ColdBox ($\Phi^{\rm in}_{\rm Th}$) to a location outside ($\Phi^{\rm out}_{\rm Th}$), was found to be:
\begin{equation}
R_{\text{BatMon}} \equiv \frac{\Phi^{\rm in}_{\rm Th}}{\Phi^{\rm out}_{\rm Th}} < \frac{2 \times 10^{5}\ [\rm cm^{-2}\ s^{-1}]}{4.40 \times 10^{6}\ [\rm cm^{-2}\ s^{-1}]} \approx 4.5 \times 10^{-2}.
\end{equation}
The simulated value $R_{\text{SIM}} = 2.3 \times 10^{-3}$ is approximately a factor $20$ lower than the measured upper limit\footnote{The thermal neutron flux inside the ColdBox ($\Phi^{\rm in}_{\rm Th}$) was measured below the detector sensitivity ($\rm 2\times10^5\ [\rm cm^{-2}\ s^{-1}]$); therefore, we can only obtain an upper limit for the measured thermal neutron flux inside the ColdBox.}. This conservative discrepancy is expected, as the simulation includes holes that are sealed in practice with Mirrobor, and the BatMon value is an upper limit. This confirms the high efficiency of the shielding design. In addition, Figure \ref{neutron_spectrum} shows that the effects of the holes (magenta, brown, and yellow lines) are not very relevant and can be neglected, particularly for energies near thermal neutron temperature. 

\section{Verification of Boron Content}
\label{sec:verification}

To verify the boron content in the borated polyethylene, a study was conducted at the Solids Analysis Laboratory (LAS) of Andres Bello University. The results obtained are presented in Table \ref{tab:composition}.

\begin{table}[h!]
\centering
\begin{tabular}{|l|l|} 
\hline
Element                               & Weight percentage  \\ \hline
B                                          & 35.73         \\  \hline
C                                          & 2.11           \\  \hline
S                                          & 2.79           \\  \hline
O                                          & 1.60           \\  \hline
H                                          & 0.07           \\  \hline
Amorphous                           & 57.69        \\  \hline
                                       & 100.00      \\ \hline
\end{tabular}
\caption{Chemical composition of borated polyethylene was calculated using the software suite Diﬀrac TOPAS 4.2 \cite{topas}.}
\label{tab:composition}
\end{table}

It is worth mentioning that in borated polyethylene, the amorphous content primarily refers to the disordered regions of the polyethylene polymer chains themselves, which are inherent to its semi-crystalline nature. This amorphous phase is crucial as it allows for the uniform dispersion and embedding of boron-based particles (like boron carbide) throughout the plastic matrix.

\section{Conclusions}
\label{sec:conclusions}

The design and construction of the ColdBox for the SND@LHC experiment successfully addresses the challenge of providing simultaneous neutron shielding and environmental insulation for the sensitive emulsion film target.

The selected configuration, combining a 5\,cm plexiglass outer layer with a 4\,cm borated polyethylene (35\% boron) inner layer, was shown through FLUKA simulations to reduce the thermal neutron flux by more than two orders of magnitude ($R_{\text{SIM}} = 2.3 \times 10^{-3}$). This simulated performance is consistent with and superior to the initial upper limit established by BatMon measurements.

Mechanically, the aluminium frame provides a robust and stable support structure, with finite element analysis confirming its integrity under operational loads. The inclusion of an aluminium heat shield ensures compliance with CERN safety regulations, and the use of Mirrobor seals enhances the overall shielding effectiveness.

The ColdBox is designed to maintain a stable internal environment by providing a sealed volume for a dedicated evaporator system. Future work will focus on monitoring the long-term performance of the ColdBox, including the validation of the temperature and humidity stability during experiment operation. A more detailed analysis of shielding efficiency could be performed with more sensitive detectors capable of measuring a broader neutron energy spectrum.

\pagebreak
\newpage
\appendix
\section{Appendix: ColdBox drawings}
\label{appendix}

\begin{figure}[H]
\centering
\includegraphics[angle=0, scale=0.48]{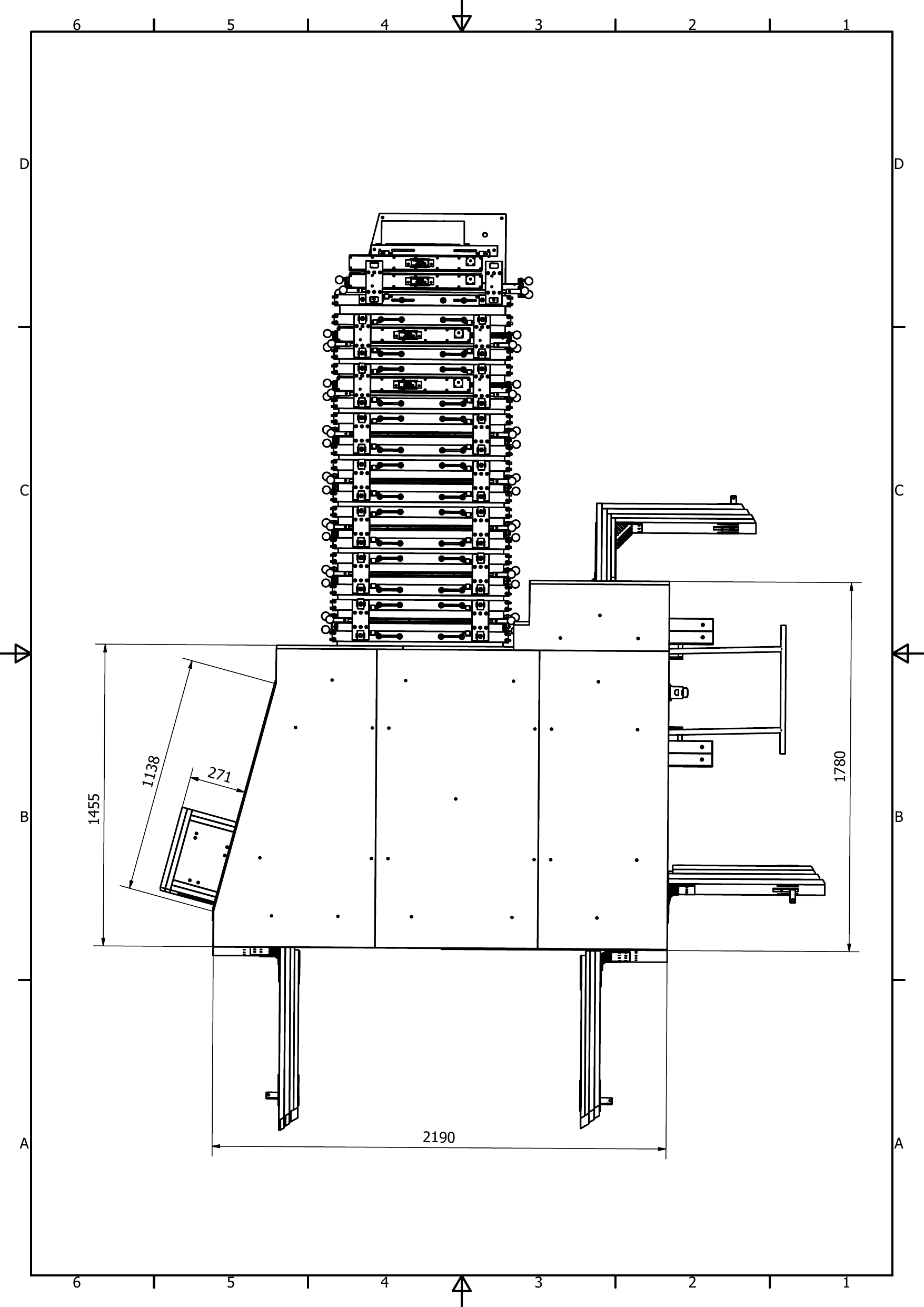}
\caption{Top view of the coldbox}
\end{figure}

\begin{figure}[H]
\centering
\includegraphics[angle=-90, scale=0.48]{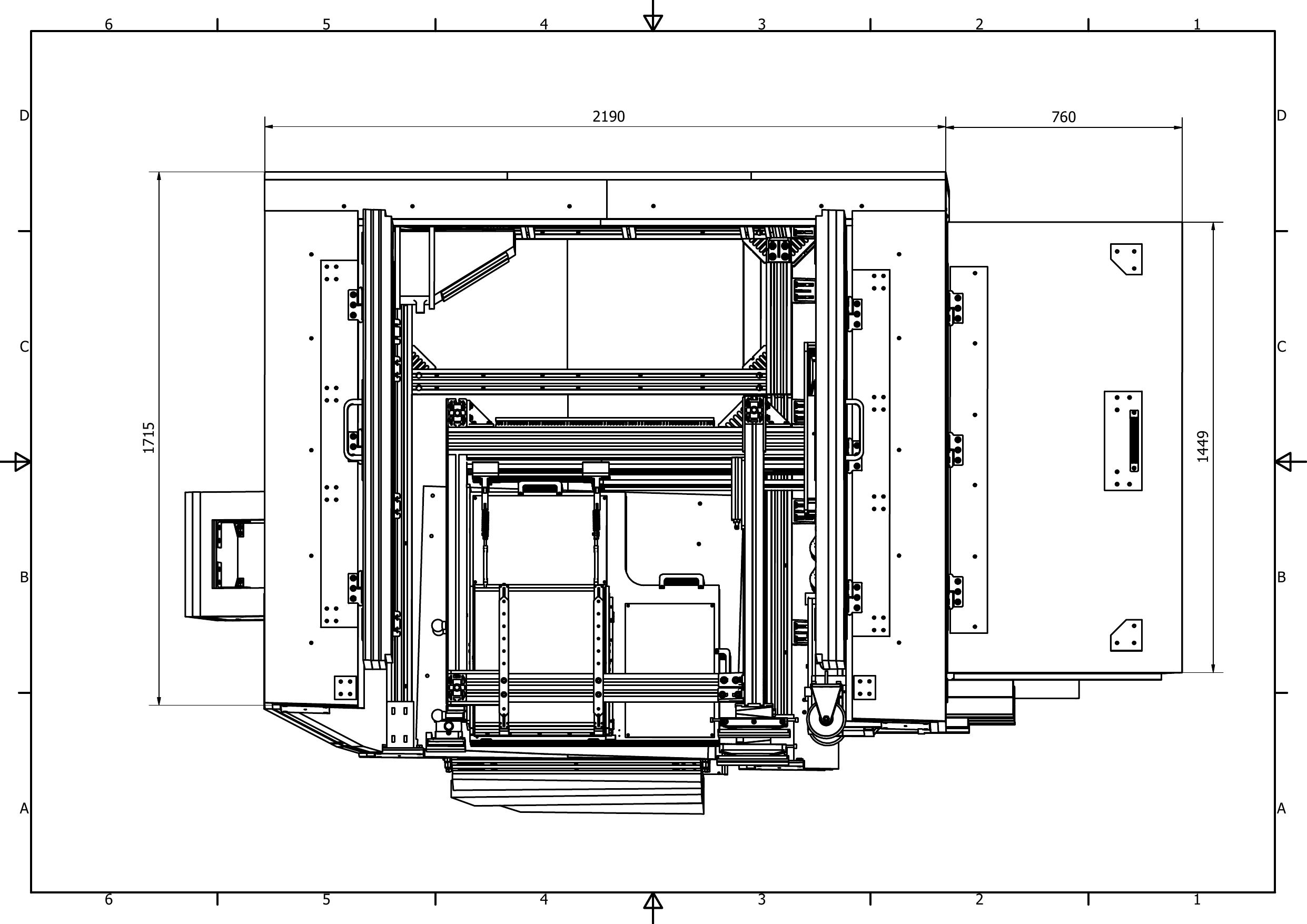}
\caption{Front view of the coldbox}
\end{figure}

\begin{figure}[H]
\centering
\includegraphics[angle=-90, scale=0.48]{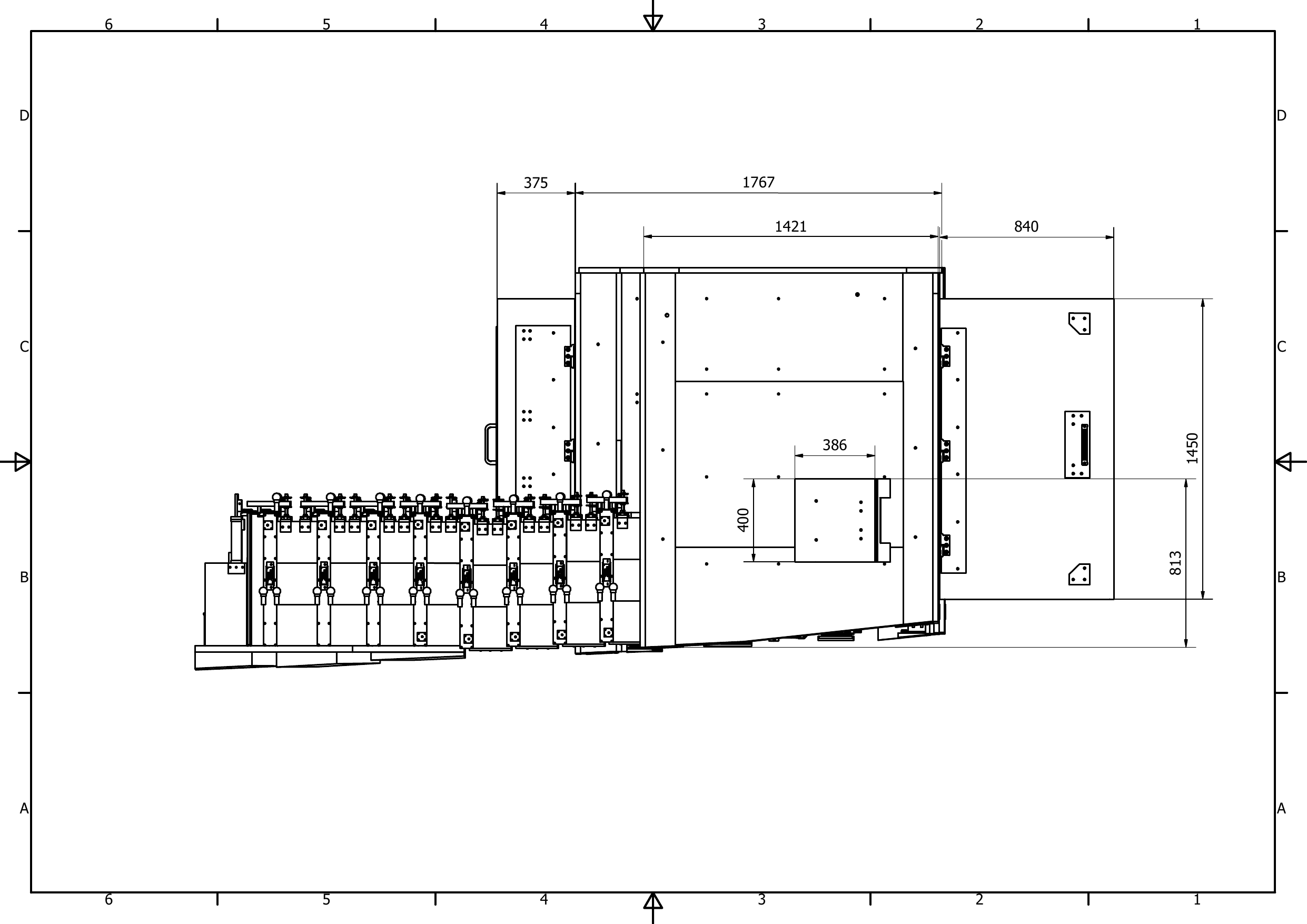}
\caption{Left view of the coldbox}
\end{figure}

\begin{figure}[H]
\centering
\includegraphics[angle=-90, scale=0.48]{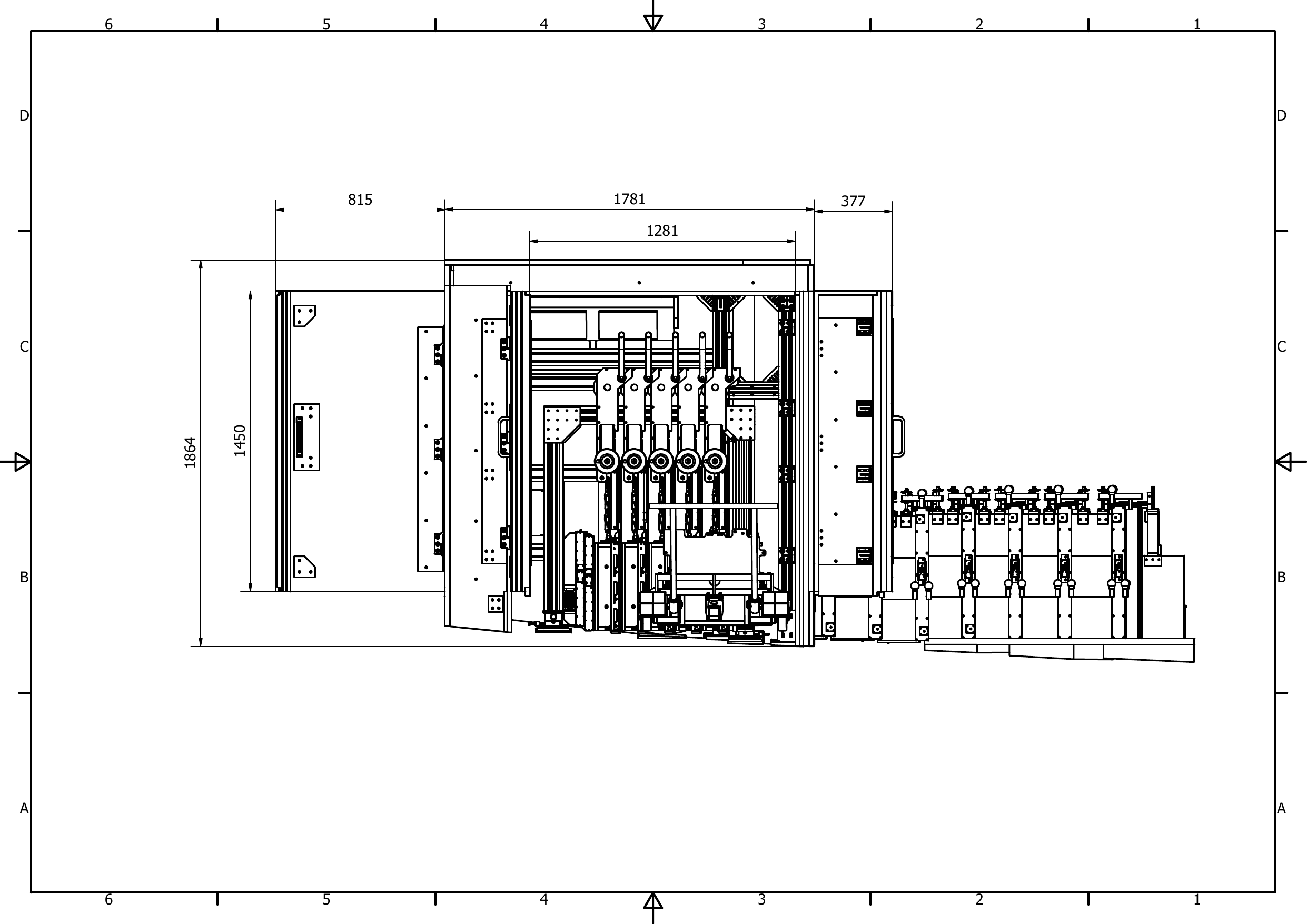}
\caption{Right view of the coldbox}
\end{figure}

\begin{figure}[H]
\centering
\includegraphics[angle=-90, scale=0.48]{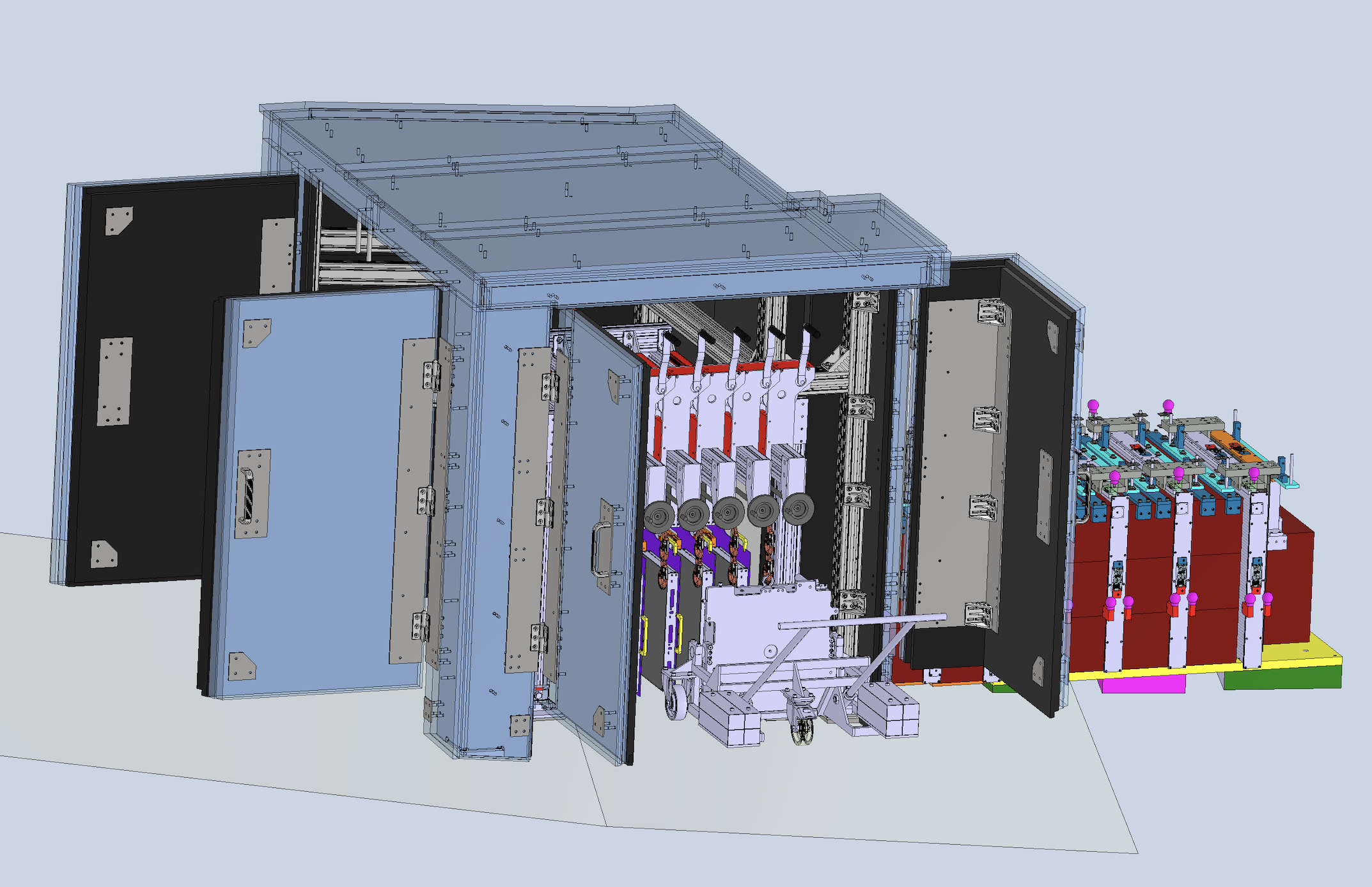}
\caption{3D view of the coldbox}
\end{figure}

\pagebreak
\section*{Acknowledgments}

We acknowledge the support for the construction and operation of the SND@LHC detector provided by the following funding agencies:  CERN;  the Bulgarian Ministry of Education and Science within the National
Roadmap for Research Infrastructures 2020–2027 (object CERN); ANID FONDECYT grants No. 3230806, No. 1240066, 1240216 and ANID  - Millenium Science Initiative Program -  $\rm{ICN}2019\_044$ (Chile); the Deutsche Forschungsgemeinschaft (DFG, ID 496466340); the Italian National Institute for Nuclear Physics (INFN); JSPS, MEXT, the~Global COE program of Nagoya University, the~Promotion
and Mutual Aid Corporation for Private Schools of Japan for Japan;
the National Research Foundation of Korea with grant numbers 2021R1A2C2011003, 2020R1A2C1099546, 2021R1F1A1061717, and
2022R1A2C100505; Fundação para a Ciência e a Tecnologia, FCT (Portugal), 
CERN/FIS-INS/0028/2021; the Swiss National Science Foundation (SNSF); TENMAK for Turkey (Grant No. 2022TENMAK(CERN) A5.H3.F2-1).
M.~Climesu, H.~Lacker and R.~Wanke are funded by the Deutsche Forschungsgemeinschaft (DFG, German Research Foundation), Project 496466340. We acknowledge the funding of individuals by Fundação para a Ciência e a Tecnologia, FCT (Portugal) with grant numbers  CEECIND/01334/2018, 
CEECINST/00032/2021 and 
PRT/BD/153351/2021. This research was financially supported by the Italian Ministry of University and Research within the Prin 2022 program.

We express our gratitude to our colleagues in the CERN accelerator departments for the excellent performance of the LHC. We thank the technical and administrative staff at CERN and at other SND@LHC institutes for their contributions to the success of the SND@LHC efforts. We thank Luis Lopes, Jakob Paul Schmidt and Maik Daniels for their help during the~construction.

\bibliographystyle{apsrev4-1}

\end{document}